\pdfoutput=1 

\documentclass[12pt]{article}
\usepackage{amsmath,graphicx,hhline}
\usepackage{geometry} 
\date{}

\begin{document}

\author{Fr\'ed\'eric Sirois$^1$ and Francesco Grilli$^2$ \\
$^1${\small Department of Electrical Engineering, Polytechnique Montr\'eal, Montr\'eal, Canada} \\
$^2${\small Institute for Technical Physics, Karlsruhe Institute of Technology, Karlsruhe, Germany}
}
\title{Potential and limits of numerical modelling for supporting the development of HTS devices}

\maketitle

{This is an author-created, un-copyedited version of an article accepted 
for publication in Superconductor Science and Technology. The publisher is not responsible 
for any errors or omissions in this version of the manuscript or any version 
derived from it. The Version of Record is available online at http://dx.doi.org/10.1088/0953-2048/28/4/043002.
}

\begin{abstract}

In this paper, we present a general review of the status of numerical modelling applied to the design of high temperature superconductor (HTS) devices. The importance of this tool is emphasized at the beginning of the paper, followed by formal definitions of the notions of \emph{models}, \emph{numerical methods} and \emph{numerical models}. The state-of-the-art models are listed, and the main limitations of existing numerical models are reported. Those limitations are shown to concern two aspects: one the one hand, the numerical performance (i.e. speed) of the methods themselves is not good enough yet; on the other hand, the availability of model file templates, material data and benchmark problems is clearly insufficient. Paths for improving those elements are provided in the paper. Besides the technical aspects of the research to be further pursued, for instance in adaptive numerical methods, most recommendations command for an increased collective effort for sharing files, data, codes and their documentation.

\end{abstract}

\section{Introduction}

The development of high temperature superconductor (HTS) devices has made significant progresses over the last few years, especially those based on second-generation (2G) HTS tapes. Prototypes are now close to commercial products, or even already commercially available, as in the case of superconducting fault current limiters and cables. Although the cost of 2G HTS tapes makes devices still expensive, steady progress in materials science and in manufacturing processes brings their price down every year. In parallel to materials improvements, it is possible to further decrease the cost of devices by minimizing the quantity of HTS material used in a given application. This is not an obvious exercise though, as HTS materials are highly nonlinear and present strong anisotropic field dependence. In most cases, only numerical methods allow relating power dissipation in HTS parts with their geometrical arrangement within the device.

In a general sense, numerical modelling in engineering is a mature discipline. Numerical methods such as the finite element method and others are well known and documented in the literature. However, each engineering application has its own specificities, which one can take advantage of in order to simplify the problem and reduce the associated computational size. This is of tremendous importance in a context of device optimization, where hundreds or even thousands of parametric simulations are often required. This is also the reason why there are dozens of variants of the established numerical methods in the literature.

Combined to this variety of applications, the materials used in the various engineering devices introduce nonlinearities that significantly affect the numerical behaviour of the problem, sometimes to the point where the mathematical theory behind the numerical method must be revisited. This is particularly true in the case of superconducting materials, where the disappearance of the resistivity at low current leads to singular behaviour of the electromagnetic fields (moving current fronts). In practice, this does not mean classical methods cannot handle the problem, but rather than blindly applying classical methods is by far sub-optimal and may lead to computation times that are much longer than required.

In this contribution, we start by reviewing briefly the modelling approches and the numerical methods that are commonly used to solve problems involving HTS materials. These problems usually represent a portion of an HTS device.
%
After reviewing the most common models and numerical methods, their limits are briefly outlined, in order to define working paths towards numerical methods that are specifically tailored for problems involving superconducting materials. In particular, we discuss the issue of necessary degrees of freedom and adaptive finite element methods that allow placing the unknowns where they are really needed. It is shown that there is still a significant potential for making numerical simulations faster and therefore developing fast optimization tools that would significantly speed up HTS devices development.

Finally, we emphasize the need for a systematic benchmarking of the various models and methods, ranging from detailed physical models to more macroscopic circuit models. Only few existing benchmarks exist at the moment, so the whole HTS community (industrialists and academics) is invited to contribute to the development of new benchmarks that are representative of the state of the art of HTS device engineering. These benchmarks will greatly help focus the R\&D effort of the numerical modelling community towards the most relevant approaches.

\section{Context and need for numerical modelling}
If we look at the status of development of HTS devices, we can see a steady progression over the last few years. Many prototypes have been built, and some commercial devices already exist. While the need to reduce the cost of 2G HTS wires is widely recognized~\cite{Melhem:2012}
and possible solutions to reach that goal proposed~\cite{Matias:PP12}, other important practical aspects need to be considered for the manufacturing of real applications, for example the cost and reliability of cooling systems or the mixture of high voltages and cryogenic temperatures, just to cite a few.

Regardless of the specific engineering problem to solve, \emph{numerical modelling} is one common tool that comes to the rescue of engineers and scientists. Not only does numerical modelling allow deepening the understanding of the behaviour of a device under various excitations, but it also allows optimizing its performance, and in turn reducing its cost by 1) making the best use of materials and components, and 2) reducing the number of prototypes during the development stage. This is particularly true for devices based on HTS materials because their highly non-linear and often anisotropic behaviour makes it difficult to intuitively find the best operational configurations.

Beyond the behaviour and the optimization of the device itself, numerical modelling can also be used to predict how a device will perform in its environment once in operation. In other words, a device rarely operates alone, but acts on its neighbourhood as one element of a system. A good example of this is a superconducting power device installed in a power system. Since the final performance of a device is conditioned by the system in which it is installed, it is of the highest importance to develop device models that are compatible with system simulators. These device models, most often expressed in terms of electric circuits, are in general simpler than those used for device optimization, for instance finite element models, but they are nonetheless essential for comparing the performance of competing technologies in a given system. An interesting discussion about the requirements for multi-scale modelling of magnetic devices (mostly electric motors in this case) is presented in \cite{Lowther:2013eg}.

In this paper we limit the discussion to numerical tools related to electromagnetic problems involving superconductors, knowing that we could also talk about many other types of physics (thermal, mechanical, etc.). The main reason for this choice is that, without the intention of minimizing the difficulties associated to other physics, we believe that electromagnetics is probably the most open modelling issue in the area of HTS materials. It will be seen that numerous choices of mathematical formulations exists in electromagnetics, and no obvious criteria allow identifying a ``best method''. Despite this deliberate narrowing of the topic, the major part of this paper applies to numerical modelling in general.

\section{Models, numerical methods and numerical models}
\subsection{Definitions and overview}

It is beyond the scope of this paper to give a detailed list of all the numerical tools for modelling HTS. These can be found in dedicated publications~\cite{Campbell:JSNM11, Mikitik:TAS13, Grilli:TAS14a}. In this section we would rather like to give an overview of the general principles of and differences between the various models and numerical methods available in the literature and applicable to HTS materials. We start with the distinction between {\it models} and {\it numerical methods}, two terms that are often interchanged in an improper way.

A {\it model} is a mathematical representation of a physical (or other) behavior, based on relevant hypothesis and simplifying assumptions. For example, in the case of superconductors, we have the power-law model (PLM) and the critical state model (CSM): the former considers flux creep, whereas the latter does not. Another example is Maxwell's equations, which provide a mathematical representation to model electromagnetic fields, but can be written in several ways depending on what matters for a specific problem, e.g. with or without the displacement current term, written in terms of time harmonics (i.e. $d/dt \rightarrow j \omega$) or directly with the time derivative terms, as required for superconductors since those materials are non-linear and their state evolve in time.

A {\it numerical method} consists of a systematic approach to 1) express a model in a discrete form, 2) generate a system of equations that approximates this model, and 3) solve the resulting system of equations. There exists many examples of numerical methods, for instance the finite element method, the finite difference method, etc. (see section \ref{num_methods} for details). The choice of a given numerical method should have no impact of the physical meaning of the solution: all numerical methods should theoretically converge towards the same result as the discretization is progressively refined. On the other hand, the choice of a particular numerical method can have a drastic impact on the computation time required to solve the model.

Finally, we define a {\it numerical model} as the \emph{combination of a \emph{model} and a \emph{numerical method}}. The model establishes the physical representativeness of the solution, whereas the numerical method defines the accuracy to which the model can be approximated, and determines in good part the computational speed. However, since the model is a trade-off between physical relevance and complexity, the choice of a model has also an implicit impact on the overall computation time.

\subsection{Choice of a model}

The \emph{quality} of a model highly relies on the ``smartness'' of the assumptions behind it. In other words, if the model is ``bad'' (in terms of physical representativeness), the solution will be ``bad'', regardless of the numerical method chosen to solve it.

Of course, the definition of ``bad'' is highly function of the modelling objectives. Typical considerations that need to be taken into account when choosing a model are:
\begin{itemize}
\item Is it necessary to simulate the whole device? Can symmetries/periodicities be used, or dimensionality of the problem be reduced (e.g. from 3-D to 2-D or 2.5-D)?
\item What level of physical representativeness is really required? Is a simple trend prediction sufficient? How accurate are the experimental results to validate the model?
\item Is it necessary to look in detail at local results (field, current density), or does one just need global quantities as output (e.g. AC losses)? What are the needs of industrialists and manufacturers vs. those of academics?
\end{itemize}

\begin{figure}[t!]
\centering
\includegraphics[width=5.5in]{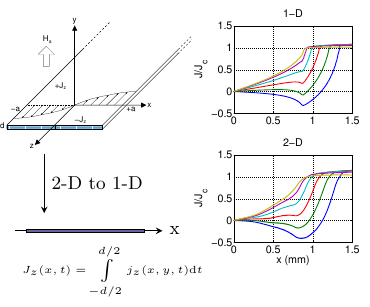}
\caption{\label{fig:2D_1D}{Schematic illustration of the reduction of a 2-D model of a thin rectangular strip to a 1-D approximation. Figures adapted from~\cite{Brambilla:SST08}.}}
\end{figure}

\begin{figure}[h!]
\centering
\includegraphics[width=5.3in]{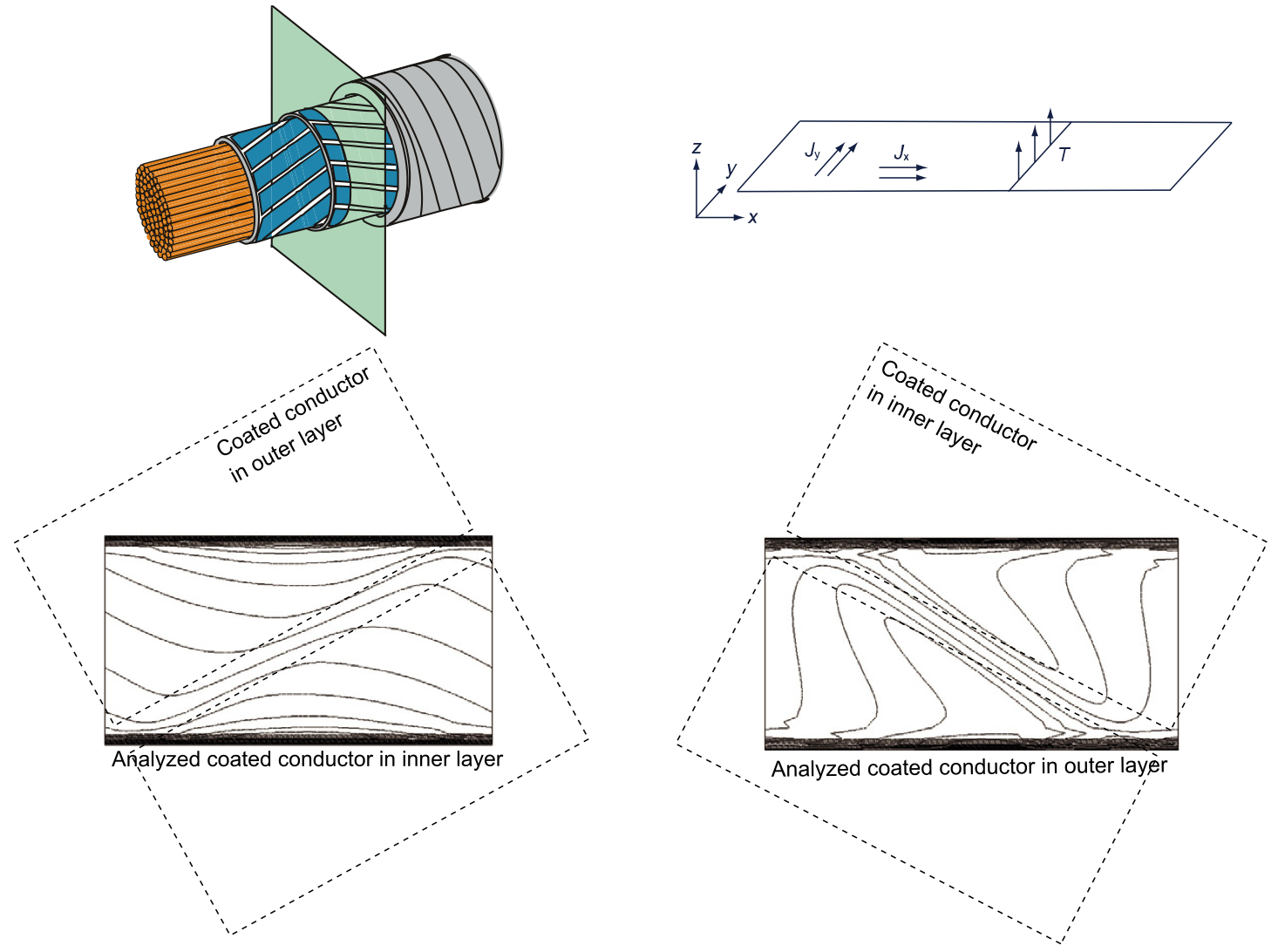}
\caption{\label{fig:Takeuchi}{3-D modelling of 2-layer spiraled cables made of 2G HTS coated conductors using a ``2.5-D'' model that considers the tapes as infinitely thin, while allowing the current to flow in the transversal direction ($y$-direction in the figure). Exemplary current patterns in the tapes of the inner and outer layers are shown. Figures adapted from~\cite{Takeuchi:SST11}.}}
\end{figure}

\begin{figure}[tbp!]
\centering
\includegraphics[width=\columnwidth]{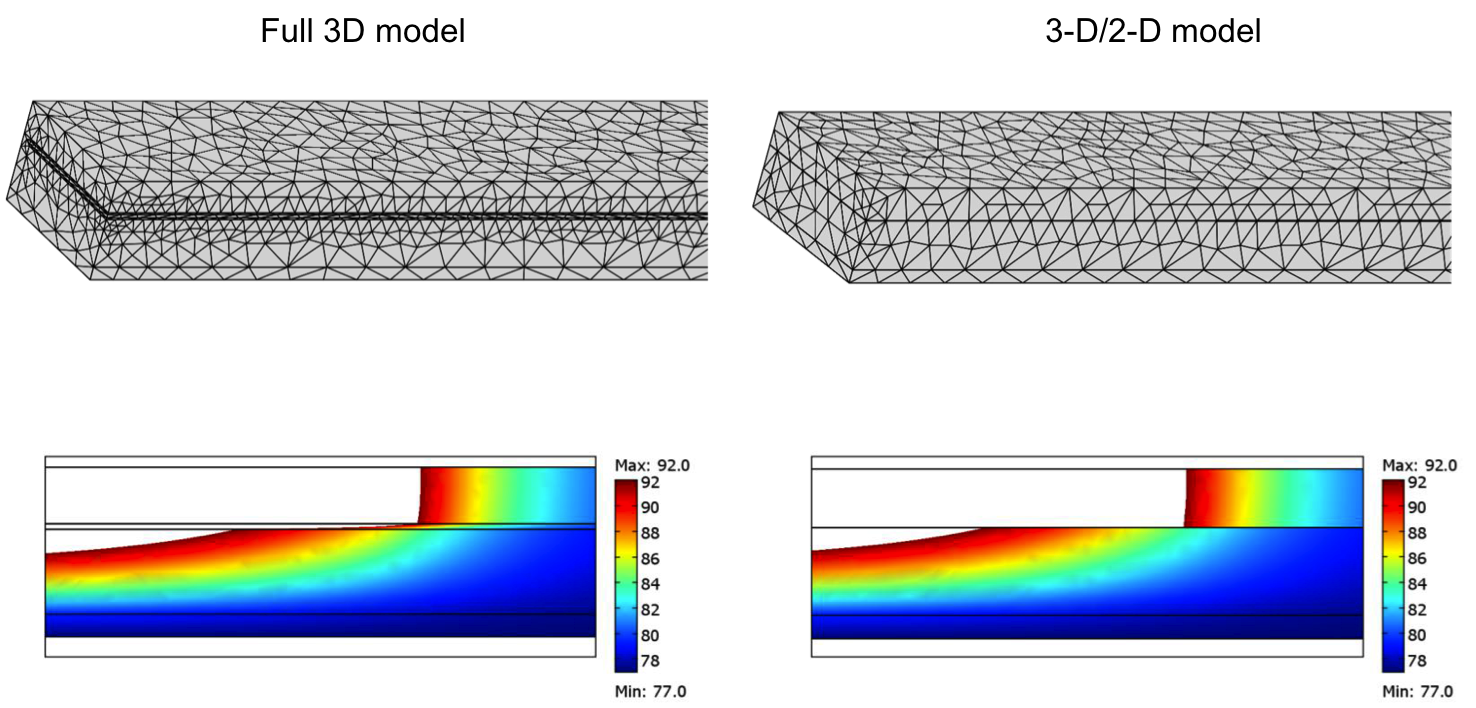}
\caption{\label{fig:Chan}{Temperature distribution a 2G HTS coated conductor computed with a full 3-D model (left) and with a thin interface condition in a mixed dimensional 3-D/2-D model (right). Figures adapted from~\cite{Chan:TAS10} and a presentation of the same authors~\cite{Lausanne_Workshop}.}}
\end{figure}

Over the years, many ``smart'' models have been developed in the HTS applied superconductivity community. The majority of models involving 2G HTS coated conductors involve dimensional reduction, for instance:
\begin{itemize} 
\item Infinitely thin film approximation of 2G HTS coated conductors~\cite{Brambilla:SST08} (see figure~\ref{fig:2D_1D});
\item 3-D modelling of 2-layer power cables or Roebel cables using a ``2.5-D'' model: infinitely thin tapes in a 3-D space~\cite{Takeuchi:SST11,Amemiya:SST14} (see figure~\ref{fig:Takeuchi});
\item Use of thin interface conditions to reproduce the thermal and electrical behaviour of buffer layers in coated conductors~\cite{Chan:TAS10} (see figure~\ref{fig:Chan}).
\end{itemize} 
\begin{figure}[tbp!]
\centering
\includegraphics[width=8 cm]{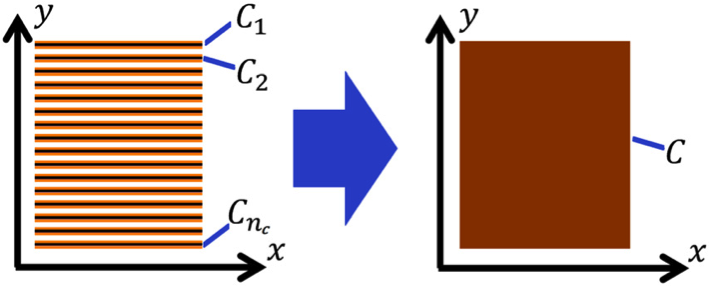}
\includegraphics[width=8 cm]{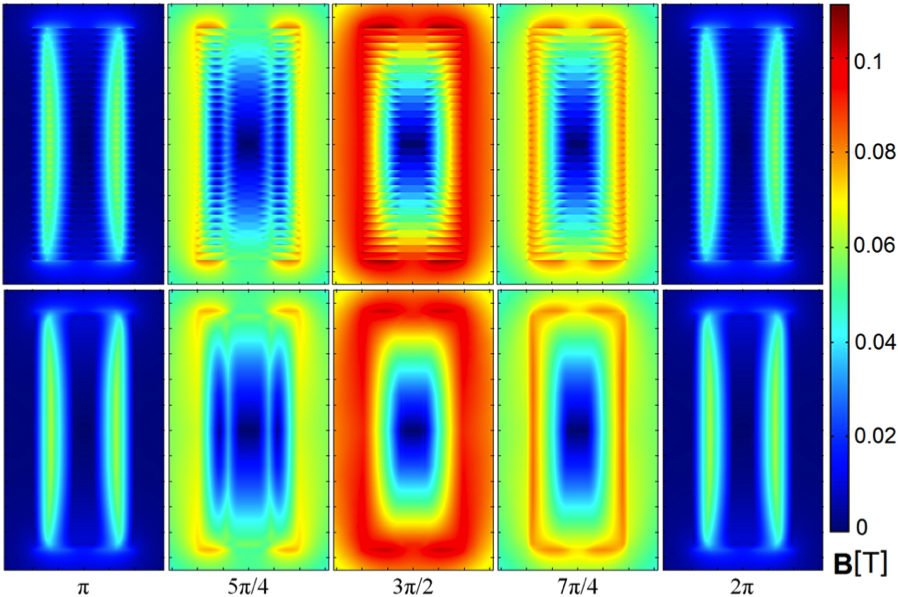}
\caption{\label{fig:homog}{Top: concept of homogenization for a stack of tapes carrying the same transport current. Bottom: magnetic flux density distribution (at different instants of the sinusoidal current) calculated for the full geometry (16 stacked tapes) and with homogenization. Figures adapted from~\cite{Zermeno:JAP13}.}}
\end{figure}
\begin{figure}[tbp!]
\centering
\includegraphics[width=6.5in]{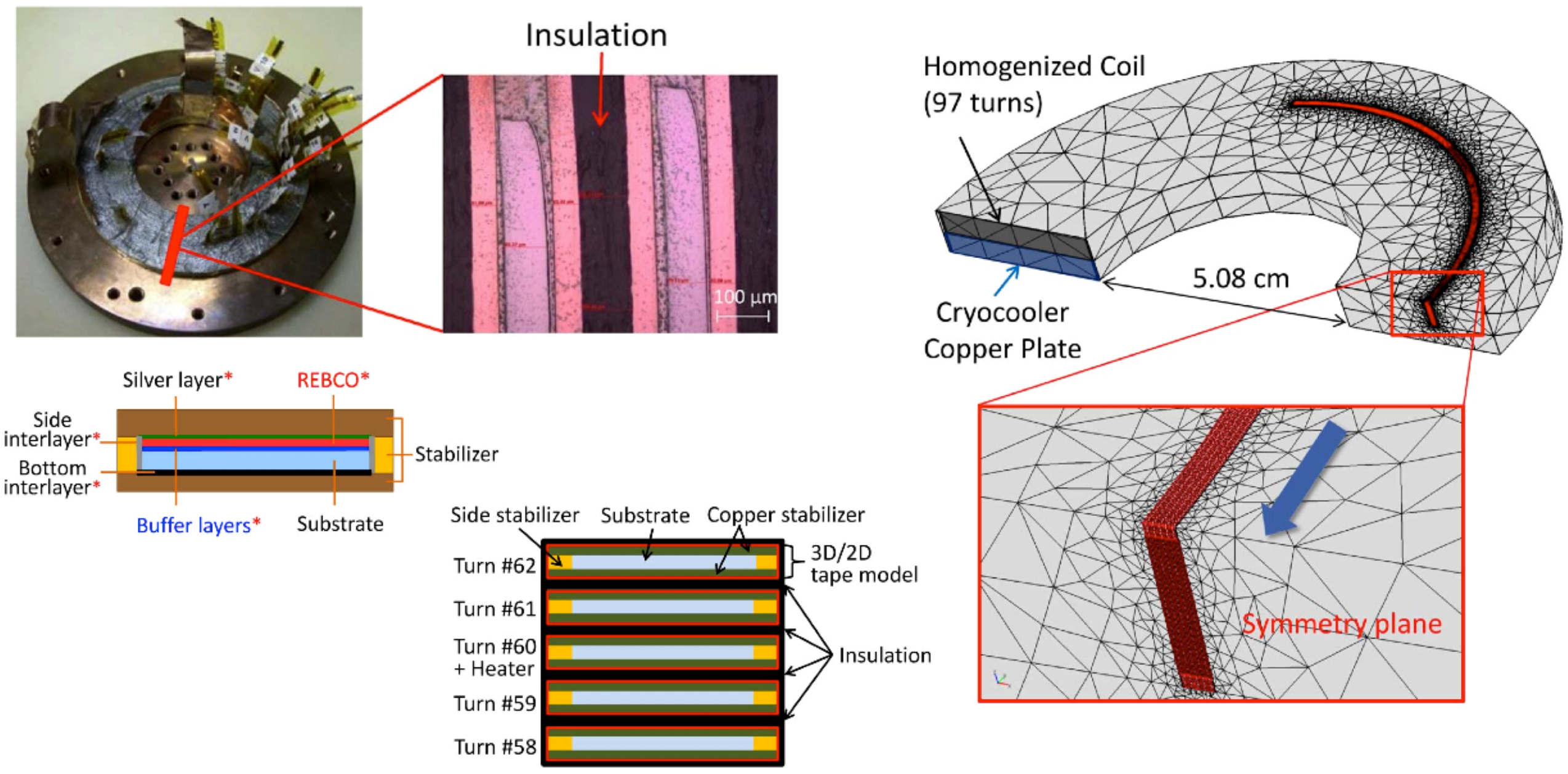}
\caption{\label{fig:Chan2}{Hierarchal, 3-D multi-scale tape model for the electro-magnetic-thermal behavior of quench in coils. Figures adapted from~\cite{Chan:TAS12}.}}
\end{figure}
In some cases, especially when many coated conductors are present and stacked on top of each other, one can homogenize the cross section of the stack, getting rid of the geometry of the individual tapes, while preserving their physical behaviour by means of special current constraints. The homogenized conductor has thus a very similar electromagnetic behaviour to that of the original tape assembly, but the solution of the problem can be obtained much more quickly, as the discrete problem is much smaller (simpler mesh, less unknowns in the problem).
The \emph{homogenization} procedure can be applied in different ways:
\begin{itemize} 
\item Direct current homogenization in a single stack of tapes~\cite{Clem:SST07, Prigozhin:SST11} and coils~\cite{Yuan:SST09} in 2-D, or even to 3-D coils~\cite{Zermeno:JAP13} (see figure~\ref{fig:homog});
\item Hierarchal current homogenization in larger assemblies of coated conductors, in order to observe local phenomena where it matters, while keeping the global system complexity to a realistic level~\cite{Chan:TAS12} (see figure~\ref{fig:Chan2}).
\end{itemize} 
Another particularly original approach is the force-displacement model developed by Campbell~\cite{Campbell:SST07}, which starts from a physical argument and reduces the critical state model to a static problem that can be solved in a single shot, without the need of time stepping algorithms.

\subsection{Formulation of a model}

Models are expressed as mathematical equations, which can be of different types, e.g. ordinary and/or partial differential equations (ODEs / PDEs), integral equations (IEs), mixture of PDEs and IEs, algebraic constraints, etc. The exact choice/combination of equations is often referred to as a \emph{formulation}.

In computational electromagnetics, which forms the basis for the analysis of superconductors, there exists many such formulations. This variety comes from the many options of state variables that one can choose to put a given problem in mathematical form~\cite{Campbell:JSNM11}. For instance, one could choose to express the problem directly in terms of the field variables (e.g. the magnetic field $\mathbf H$ and/or the electric field $\mathbf E$), or in terms of potentials (${\mathbf A}-V$, ${\mathbf T}-\Omega$, and many variants of these).

While the choice of a formulation is in principle arbitrary, certain models are more naturally expressed in terms of a specific formulation. As an example, Campbell's force-displacement method~\cite{Campbell:SST07} is naturally written in terms of the magnetic vector potential ${\mathbf A}$, so not using the ${\mathbf A}-V$ formulation in this case would be cumbersome.

Another example that may dictate the use of a potential formulation is when one needs to couple an electromagnetic model with an electric circuit \cite{Meunier:TMAG03, LeFloch:TMAG03}. This coupling requires computing global quantities, such as flux linkages or inductances, which is straightforward to do with a potential formulation, but harder when working in terms of field variables. On the other hand, it is often easier to write the model in terms of field variables, as it makes everything more intuitive to interpret.

In general, there is no clear choice when choosing a formulation, as it depends on many factors, including some of numerical nature. Indeed, some formulations will generate more unknowns in the numerical problem, but may in turn be more stable (numerically speaking) than a less memory demanding formulation.

\subsection{Choice of a numerical method}
\label{num_methods}

A given model can be solved by different numerical methods. Once again, it is out of the scope of this paper to review all existing numerical methods, but table~\ref{table:num_methods} gives a brief summary of those encountered in modelling devices involving HTS materials. Acronyms used in table~\ref{table:num_methods} are common, but not necessarily the only ones used in literature.

Each method typically has many variants, which we shall not refer to in this paper. However, there is one common point among all numerical methods: their role is to provide a \emph{discrete approximation of the exact solution} of the model, whichever formulation one chooses to describe the original problem. This requires to discretize a priori the geometry of the problem, either as a grid or as a mesh, depending of the method to be used afterwards. The dimension of this grid/mesh, i.e. number of nodes or edges in it, is directly related to the number of \emph{degrees of freedom} (DOFs, or more simply, \emph{unknowns}) of the problem. The number of DOFs obviously has a major impact on the computation time and memory requirements.

\begin{table}[bp]
\caption{Summary of families of numerical methods and their main features. Acronyms stand for: Finite Differences (FD), Finite Volumes (FV), Finite Elements (FE), Method of Moments (MoM), Integral Equations (IE), Boundary Elements (BE), Minimization of Energy (MoE), Minimum Magnetic Energy Variation (MMEV). }
\begin{center}
\begin{tabular}{|c||c|c|c|c||c|}
\hline
Form of 	& Strong form	& Strong form	& Weak form	& Energy	 	& Matrix \\
equations	&				& (averaged) 	&				&  functional	& struct.\\
\hhline{|=|=|=|=|=||=|}
Differential	& FD	& FV	&	FE	 	& 			& Sparse \\
\hline
Integral 		& MoM	&  		&	IE, BE$^*$	&	MoE/MMEV 	& Full \\
\hhline{|=|=|=|=|=||=|}
Discretization 	& Grid	& Mesh		& Mesh		& Mesh \\
\hhline{|-|-|-|-|-||}
Type of 	& Point			& Point				& Weighted		&  Error mini-\\
method		& collocation	& collocation		& residuals		&  mization \\
\hhline{|-|-|-|-|-||}
\end{tabular}
\end{center}
\vspace{-1ex}
\hspace{19ex}
$^*$ \small{Hardly applicable to nonlinear problems.}
\label{table:num_methods}
\end{table}%

\subsubsection{Solution of the model in its strong form}

A pure \emph{strong form} solution means that one solves a pointwise version of the differential or integral equations defining the model \emph{exactly} at every grid point (for FD and MoM). These types of methods are often referred to as \emph{point collocation} methods, as one freely interpolates the pointwise solution between each point of the grid to form a complete solution in space.
A widely used strong form solution in applied superconductivity is the so-called ``Brandt method'', which is in fact a method of moments (MoM) based on integral equations. The method is well described in \cite{Brandt:1996ez}, and has been used by many authors over the years.

It is good to mention that, because integral equations (involving Green functions, which are defined everywhere in space) are by nature smoother than differential equations (defined only locally), approximation errors naturally tend to be better distributed with the MoM than with FD. This is not a strict rule though, as Green functions may involve singularities that can lead to very large errors when they are not treated properly.

Although strong form methods are very intuitive and among the easiest numerical methods to program, they have a very important drawback: they do not provide good control on the approximation error over the domain. In addition, for complex geometric shapes, a simple grid discretization often provides a very poor approximation of the boundaries of the different geometric objects.

One particularly important case of \emph{strong form} methods is FV. This method is somewhat midway between FD and FE (see next section), since the equations to solve involve integrals of fluxes (e.g. electric currents) crossing the boundary of every elements (control volumes) of the mesh. This integration procedure recalls the weak form encountered in FE, although equations are still solved in a pointwise manner, so it remains in the category of point collocation methods. The major advantage of this method is that it implicitly makes the flux variable divergence-free by equating the flux integrals on every common edge (in 2-D) of face (in 3-D) of mesh elements. Although this method is widely used in fluid mechanics for incompressible flow models, it has scarcely been applied to electromagnetic problems \cite{Kameni:2010cq}. 

In brief, strong form methods, despite their simplicity of implementation, often present limitations when it comes to complex geometries or large scale problems. They are interesting approches to quickly start an investigation, but in general, it seems preferable to opt for a \emph{weak form} or \emph{energy functionnal} approach.

\subsubsection{Solution of a weak form of the model}
Instead of solving very strictly the equations defining the model, an interesting strategy is to multiply the original equations by weighting functions (often called test functions), and then integrate those modified equations over the problem domain. The modified equations are said to be written in \emph{weak form}. Although this modifies the original system of equations, it also brings up the option of weighting the numerical error over the whole domain, which is impossible with point collocation methods. In other words, the numerical solutions one can find in this way do not provide optimal pointwise accuracy, but they minimize the error on the overall domain, according to the chosen weighting pattern. This class of method is called \emph{weighted residuals}, and it entirely relies on functional analysis in its mathematical theory.

The basic principle of all weighted residual methods is to define \emph{shape functions} that are used to approximate in a piecewise manner the continuous solution of a problem on a discrete mesh. One usually refers to FE (finite elements) to describe this method, although the term is generally reserved for the case where one starts from PDEs to write the weak form. When dealing with integral equations, it is possible to apply the same principle to get an equation system. In this paper, we use the IE acronym to distinguish it from FE. A particular case of IE is BE (boundary elements), which is expressed exactly as IE, except that we apply a dimensionality reduction at the very beginning of the procedure in order to reduce the mesh complexity (from 3-D to 2-D, or from 2-D to 1-D). However, BE do not apply well to nonlinear problems, such as those with superconductors, and we shall not discuss this method here.

Among all numerical methods used in engineering for design and analysis purposes, FE is by far the most popular one. This is due to the generality of its formulation, its ability to model geometries of complex shapes in any dimension (1-D, 2-D, 3-D), and its relative simplicity of use, especially through many commercial software packages. This observation also holds true in the applied superconductivity community.

\subsubsection{Minimization of an energy functional}
One can define a third family of numerical method based on the minimization of an \emph{energy functional}. This approach is very intuitive, since it consists in defining a functional that relates the total energy of a system (or a variation of energy with respect to some initial conditions) with the variables that define the state of this system, e.g. potentials, field variables, source terms, etc. 
It allows more freedom in the way one chooses the shape functions used to approximate the solution. It also allows solving classes of problems that would be quite hard to solve otherwise, namely the critical state problem, which is singular in its pure form, and therefore can only be approximated to some extent when using a classical electromagnetic formulation. 
Although at first sight this approach requires less mathematical formalism than strictly applying the finite element method, the process of minimizing a functional in order to obtain a well-posed discrete equation system requires good skills in functional analysis and optimization algorithms.
Also, since it is generally based on integral equations, its use in 3-D problems is often too time consuming to be useful.

In the HTS community, the use of the MoE method was first introduced by Bossavit~\cite{Bossavit:TMAG94}, then formalized by Prigozhin~\cite{Prigozhin:JCP96}, who provided a systematic approach to solve the $J$ distribution in HTS domains based on the critical state model. A variant of the method, called MMEV, was later introduced by Sanchez~\cite{Sanchez:PRB01} and generalized by Pardo~\cite{Pardo:SST07} to include current constraints.

\subsubsection{Differential vs. integral methods}
It is interesting to look at numerical methods according to the way basic equations are written. These can be expressed either as differential or integral equations (before any averaging, transformation to weak form, etc.). This very fundamental choice has a huge impact on the numerical performance of the method.

In the case of differential equations, the solution over one element is determined solely by the values at the boundary of this element. Information is thus propagated from element to element through their common boundaries, and eventually, it is the boundary conditions imposed at the periphery of the model that establishes the complete solution. The consequence of this is that the connectivity between the degrees of freedom is only function of the nodes (in case of nodal base functions) or edges (in case of vector base functions) shared by neighbouring elements, leading to sparse matrix patterns, and a more or less linear relationship between the number of DOFs and the memory requirements.

In the case of integral equations, things are radically different: in this case, all regions containing source terms contribute to the solution on other elements. For instance, one can think about the Biot-Savart formula, which allows computing field values anywhere in space, even if the source term is a punctual current. This means that each degree of freedom in the problem is related to the others, and the resulting matrix pattern is a full one. The consequence is that the memory requirements grows at least with the square of the number of DOFs. In counterpart, there is no more need for boundary conditions anywhere in the model, but it is not so easy in practice to take advantage of this fact.

\begin{figure}[tbp!]
\centering
\includegraphics[width=8cm]{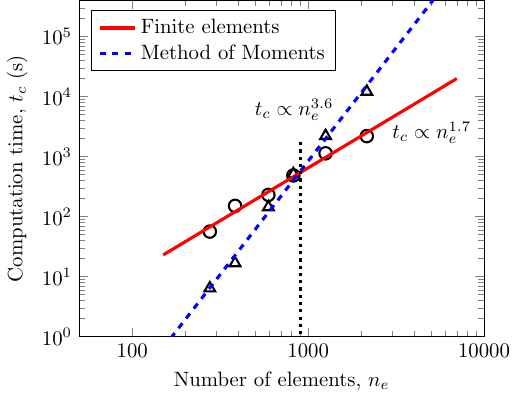}
\caption{\label{fig:tc_dif_int}{Example of computational time growth of a differential method (FE) and an integral method (MoM) as the number of elements in a reference conductor. Figure adapted from~\cite{Sirois:TAS09}.}}
\end{figure}

A particular aspect of computational electromagnetism is that air regions, which contain no source term, still need to be meshed when using differential methods, whereas they can be discarded when using integral methods. Overall, the resulting requirement in term of memory and DOFs depends on the proportion of mesh that can be avoided in a given problem. In general tough, large problems scale better with differential methods, and as soon as the number of DOFs exceeds a few thousands (which is very common in models of a practical device), differential methods seem to be the only practical way in order to perform numerical simulations, as shown in figure~\ref{fig:tc_dif_int}~\cite{Sirois:TAS09}.

\subsection{Computational performance of numerical methods}

It is extremely difficult to provide a general relationship relating the computation time ($t_c$) of a numerical method to the number of DOFs ($N$) in a given problem: the relationship would typically take the form $t_c=a+N^q$, where $q>1$ is the only thing one can be sure of ($q>3$ is common with integral methods). The exact relationship depends on the algorithm chosen to solve the problem, the need for Newton iterations (if problem is nonlinear), the conditioning of the matrix, which may require pre-processing of the linear system, etc. Also, if the problem is very large, limitations in computer memory may dictate the use of an iterative solver, which may require many iterations to converge. Therefore, here we limit ourselves to ball park figures of solution times obtained for typical electromagnetic problems involving HTS devices (e.g. AC losses vs. transport current) and solved by FE, as shown in table~\ref{tab:tcomp}.

\begin{table}[t!]
\renewcommand{\arraystretch}{1.3}
\centering
\caption{Typical numbers of DOFs and solution time of time-dependent FEM models.\label{tab:tcomp}}
\begin{tabular}{lll}
Dimension~  & \# of DOFs & Solution time \\ \hline
2-D & 1,000 to $>$100,000 & Minutes to hours \\
3-D & 100,000 to $>$1,000,000~~  & Hours to days \\
 \end{tabular}
\end{table}

These computational times, although relatively long in general, might be acceptable for helping understand the behaviour of the simulated device, but they are still too long to be used for optimizing devices, which typically require hundreds or even thousands of simulations.

\subsection{Is there any optimal choice for modelling HTS devices?}

Many numerical models for simulating the electromagnetic behaviour of superconductors have been developed in the past, but they have been seldom compared. In this section, we briefly summarize the work of the few authors who have done such comparisons.

Sykulski {\it et al.}~\cite{Sykulski:TM97} modelled AC losses in HTS as a highly non-linear diffusion process (1-D only). They presented formulations in terms of both H and E and they argued, using dimensional analysis, that working with E is numerically more efficient. Results calculated using a finite-difference scheme were shown.

Vinot {\it et al.}~\cite{Vinot:TMAG00} compared the $\mathbf{A}-V$, $\mathbf{T}-\Omega$ and $\mathbf{E}$ formulations within the FEM software package Flux3D (now simply called Flux~\cite{Company:Flux}). They found that the $\mathbf{T}-\Omega$ is the most efficient, although the $\mathbf{E}$ formulation can be used as well, provided that the index $n$ in the power-law resistivity of the superconductor is not too high ($n<20$); on the other hand, the $\mathbf{A}$ formulation proved to be not very stable.
Always in the framework of the Flux software, a table presenting a rough assessment of the $\mathbf{A}-V$ and $\mathbf{T}-\Omega$ formulations for 2-D and 3-D problems is given in~\cite{Grilli:TAS05a}, although no detailed comparison is made in the paper.

In~\cite{Sirois:JPCS08} the authors compare speed and accuracy of two FEM software packages for the 2-D simulation of a superconductor with rectangular cross-section in a variety of working conditions. In particular, they compare the $\mathbf{A}-V$ formulation with second-order nodal elements implemented in FLUX and the  $\mathbf{H}$ formulation with first-order edge elements implemented in COMSOL. It was shown that both programs gave very similar results in terms of AC losses when the current or the applied field are sufficiently large; however, some discrepancy in the result was  observed for low fields and current. The adaptive time integration algorithm used by COMSOL, based on \cite{Brenan:1996un} seems much more efficient than that used by Flux, except at large magnetic fields.

More recently, Lahtinen {\it et al.}~\cite{Lahtinen:2012bp} compared three self-programmed 2-D electromagnetic formulations ($\mathbf{H}$, $\mathbf{T}-\phi$ and, $A-V-J$) that they programmed themselves in a common computational environment. Their analysis focused on many aspects of the problem, and different methods performed differently depending on the context. Nevertheless, they concluded that the $A-V-J$ formulation seemed to be the fastest one when the most general cases were considered (combination of AC currents and AC fields).

The few above-mentioned references are not numerous enough to conclude anything about the best numerical models or formulations for a given use. Most comparison were realized in a different context (different problems, dimensionality, constraints, etc.), which might explain why conclusions differ so much, in particular regarding the choice of the most performant formulation. This indicates a lack in \emph{benchmarking}, i.e. well defined reference problems that can be used to compare the numerical performances of various codes, either commercial or home-made, based on different formulations and numerical methods, and potentially implemented on different platforms. Although benchmarks do not directly provide solutions to the modelling difficulties, they certainly help focus the research efforts and make faster progress, as demonstrated over the years in the T.E.A.M. project initiated about 30 years ago by the magnetic modelling community~\cite{Website:TEAM}.

As far as the authors know, a comprehensive review of the various numerical models developed for modelling HTS devices does not exist, but a nice summary is given in~\cite{Campbell:JSNM11}, in which a short section dedicated to the comparison of different numerical methods also reveals the lack of a proper benchmarking.

\subsection{Availability of numerical models for modelling HTS devices}

Besides the performance of numerical models themselves, an important aspect to consider is their availability. Simulators for HTS devices can be found both as home-made proprietary codes (\cite{Pecher:2003wm},~\cite{Pellikka:2013ha} and many others) as well as commercial codes
(COMSOL, Flux, ANSYS, JMAG, MagNet, FlexPDE, etc.).
There are pros and cons in both cases:
\begin{itemize}
\item Commercial codes are better supported, but they are generally expensive to purchase and maintain over time;
\item Home-made codes are more flexible than commercial codes, which is important in a field in development, but they are generally poorly documented.
\end{itemize}

Besides the simulation tools themselves, it is worth noting that typical model files are generally difficult to find. Typically, everybody has to create his/her own models from scratch, as no file template is publicly available. This is true both for finite element models and circuit models for power system simulators (i.e. no library of ``HTS devices'' exists yet).

These problems  lead to a substantial bottleneck: because of the little access to model files, and to a lesser extent, to the simulators, the modelling of HTS devices remains a specialized topic, mostly accessible to graduate students or researchers, and HTS devices remain an obscure object for most manufacturers and power utilities.

\section{Accuracy of current numerical models}

A fundamental part of the numerical model is the material model itself. In problems involving superconductors, this generally means that a good model for $J_c$  is required, which can take various forms. For instance, in the case of quench problems, one models the temperature dependence $J_c(T)$ between the operating temperature and room temperature, which is not always easy to measure, especially at high $J$ values. However, this is more an experimental characterization problem than a modelling one~\cite{Sirois:2009tn}.

In the case of applications not involving quench, most models are used to compute AC losses, and the simplest materials models are based on a constant value for $J_c$, which is usually determined by dividing the measured critical current by the cross-section of the wire/tape. However, wires and tapes used in real applications, such as coils and cables, are exposed to a magnetic induction $B$ that is generally too high to be negligible. An accurate description of the superconductor's properties should therefore include a $J_c(B)$ dependence, such as the simple one proposed by Kim for hard superconductors in the 1960s~\cite{Kim:PRL62}
\begin{equation}
J_c=\frac{J_{c0}}{1+\frac{|B|}{B_0}},
\end{equation}
where $J_{c0}$ and $B_0$ are characteristic parameters.

Increasingly complicated $J_c(B)$ formulas have been proposed, in order to take into account the peculiarities of the in-field behaviour, most notably the  dependence of $J_c$ on the orientation of the magnetic field (see for example~\cite{Stavrev:TAS02}).  For this purpose, a commonly used formula  is the elliptical field-dependence 
\begin{equation}\label{eq:JcB}
J_c (B_\perp  ,B_\parallel ) = \frac{{J_{c0} }}{{\left( {1 + \frac{{\sqrt {B_ \perp ^2  + {B_\parallel^2 }/{\gamma ^2 }} }}{{B_0 }}} \right)^\beta  }}.
\end{equation}
where $J_c$ depends on the  components of the magnetic field parallel and perpendicular to the flat face of the tape by means of the anisotropy parameter $\gamma$  (with $\gamma>1$). The cut-off value at low fields and the rapidity of the $J_c$ reduction are given by $B_0$ and $\beta$, respectively. 
With the introduction of artificial pinning centers in the superconductor, the simple description given by the four-parameter expression~(\ref{eq:JcB}) is no longer adequate; more elaborated expressions are necessary to describe the experimentally observed behavior of the samples~\cite{Pardo:SST11}.

Generally speaking, however, once the $J_c(B)$ form is known, its introduction into numerical models does not present particular difficulties, since the magnetic field variable is directly available (as in the $H$-formulation) or easily obtainable from the utilized state variables (as in potential formulations). The same comment holds true for the $J_c(T)$ relationship.

Another factor that is sometimes necessary to take into account for having an accurate model, is the spatial variation of $J_c$ inside the superconducting tape, most notably along the width of superconducting tapes. This can be the consequence of the manufacturing process, see for example \cite{Grasso:PhysC95} and \cite{Amemiya:PhysC06b} for first and second generation HTS tapes. Several techniques can be used to extract this lateral dependence.
The use of a position-dependent $J_c$ for AC loss calculations can often give a better agreement with experimental data~\cite{Grilli:TAS07a, Solovyov:SST13}.
A position-dependent $J_c$ can also be used to simulate the presence of defects~\cite{Grilli:Cryo13, Lacroix:SST14}.
Similarly to the $J_c(B)$ dependence, once the form for the spatial variation of $J_c$ is chosen, its implementation in the model does not present particular obstacles.

\begin{figure}[b!]
\centering
\includegraphics[width=8 cm]{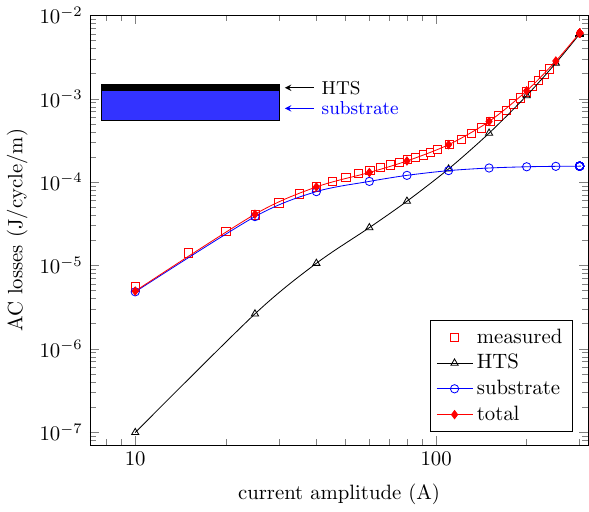}
\caption{\label{fig:doan_new}{FEM and experimental results for the AC loss components in a 1 cm-wide RABiTS YBCO tape ($I_c$ = 330 A) carrying an AC transport current~\cite{Nguyen:SST10}. Here the substrate exhibits ferromagnetic hysteretic losses that were added to the HTS losses in order to reproduce accurately the total losses of the tape.}}
\end{figure}

\begin{figure}[tb!]
\centering
\includegraphics[width=8 cm]{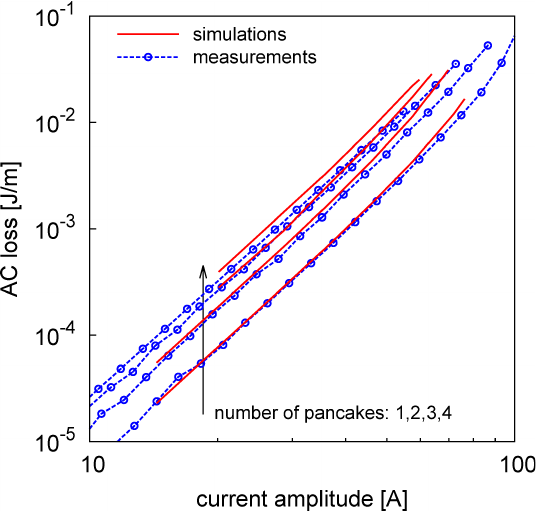}
\caption{\label{fig:pardo_pancake}{Calculated and measured AC losses in stacked pancakes of coils made of HTS coated conductors. The number of pancakes are 1, 2, 3 and 4 in the arrow direction. The agreement with experimental data is very good. Figure from~\cite{Pardo:SST12b}.}}
\end{figure}

\begin{figure}[tb!]
\centering
\includegraphics[width=8 cm]{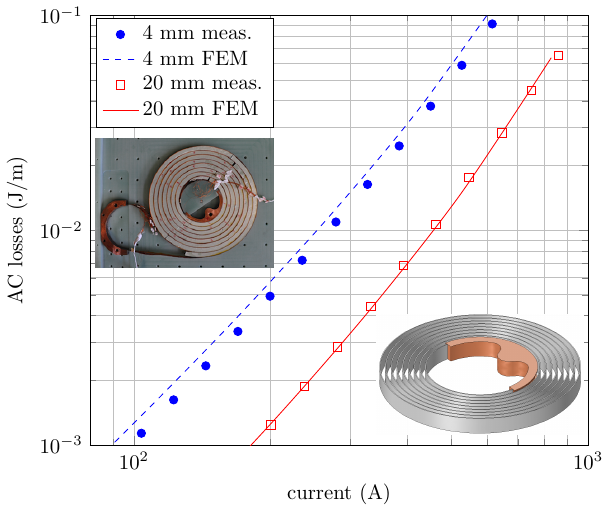}
\caption{\label{fig:QCoil_4and20mm_plot}{Measured and calculated transport AC losses for Roebel cable coils with separation between the turns of 4 and 20 mm separation between the turns. The simulation results include an anisotropic angular dependence of $J_c$ and the 3-D simulation of the copper contact for current injection, which influences the measured losses~\cite{Grilli:TAS14b}.}}
\end{figure}

Figures~\ref{fig:doan_new}, \ref{fig:QCoil_4and20mm_plot} and \ref{fig:pardo_pancake} show practical cases of simulations involving HTS models that have been compared to measurements. In all cases, the agreement with experimental data is very good. This shows that, as long as good material models are available and that one chooses a numerical model that considers the relevant physical properties of the problem, it is possible to achieve very good accuracy in the predictions, generally within 20\% of the experimental values.

It is worth mentioning that, although the implementation of complicated material properties in simulators is not difficult as such, obtaining those materials properties is not always easy, especially if one requires a model that depends on various physical parameters (field amplitude and angle, temperature, spatial inhomogeneity). A good material characterization may require a lot of a priori dedicated experiments, thus a lot of time and resources. An increased sharing of materials data would therefore help a lot modellers to achieve their goals.

Besides the cases described above, there still remains some challenges in materials modelling though, in particular with 3-D geometries, for instance when flux cutting occurs~\cite{Campbell:JSNM11}.
In an attempt to solve this problem, Bad{\'\i}a-Maj{\'o}s and L{\'o}pez have developed a general framework for handling the $E-J$ characteristic at a macroscopic level in a 3-D environment, without having to consider explicitly the microscopic effects related to vortex dynamics, but which still reproduce all relevant experimental phenomena such as flux cutting, flux flow and magnetically anisotropic critical current densities \cite{BadiaMajos:2012dt}.
Another case of interest is when the materials operate in self-field conditions and one needs to extract a truly local $J_c(B)$ dependence, whereas the experiments only provide a $J_c$ as a function of the applied field $H_a$ (see for example~\cite{Grilli:TAS14c} and references therein).

\section{Paths towards improvement}

Based on the above review of the situation, we can summarize the situation of modelling in applied superconductivity as follows:
\begin{itemize}
\item numerous numerical models exists for modelling HTS devices; however, material data are not widely available;
\item numerical simulators for HTS devices are either costly or poorly documented/supported;
\item model files are even less available and require a lot of expertise to build from scratch, which prevents potential users to simulate HTS devices in their context of application;
\item the calculation times required for running models is still too long to be of practical use in device optimization;
\item the accuracy of the solutions obtained by numerical simulation can be very good, as long as one has access to good material models.
\end{itemize}

Therefore, besides the improvement of the models themselves, this essentially defines two paths for improvement, i.e.
\begin{enumerate}
\item{improvement of the computational performances of the numerical simulators;}
\item{improvement of the availability and easiness-of-use of simulation tools (including model files and materials data).}
\end{enumerate}

Each path for improvement is discussed in more detail in the two sections that follow.

\subsection{Computational performance of numerical simulators}

As mentioned earlier in this article, there are two ingredients required for devising a good numerical simulator: 1) a good model, and 2) a suitable numerical method to solve this model. The choice of a proper model highly depends on the objective sought. New models will certainly be proposed over the years, and there is plenty of room for smart modelling ideas in the applied superconductivity community (new ideas for dimensional reductions, symmetries, mathematical formulations, etc.).

Once the model is chosen, the main aspect that governs the computational performances is the choice of the numerical method used to solve the discretized version of the model. From a practical point of view, for a given level of accuracy (determined by the user), the ``optimal'' numerical method should be able to minimize the number of degrees of freedom (DOFs) in the problem. Since the number of DOFs (or unknowns) is the main factor affecting the simulation time, minimizing it is the key issue for achieving fast simulations.

\begin{figure}[t!]
\centering
\includegraphics[width= 6 cm ]{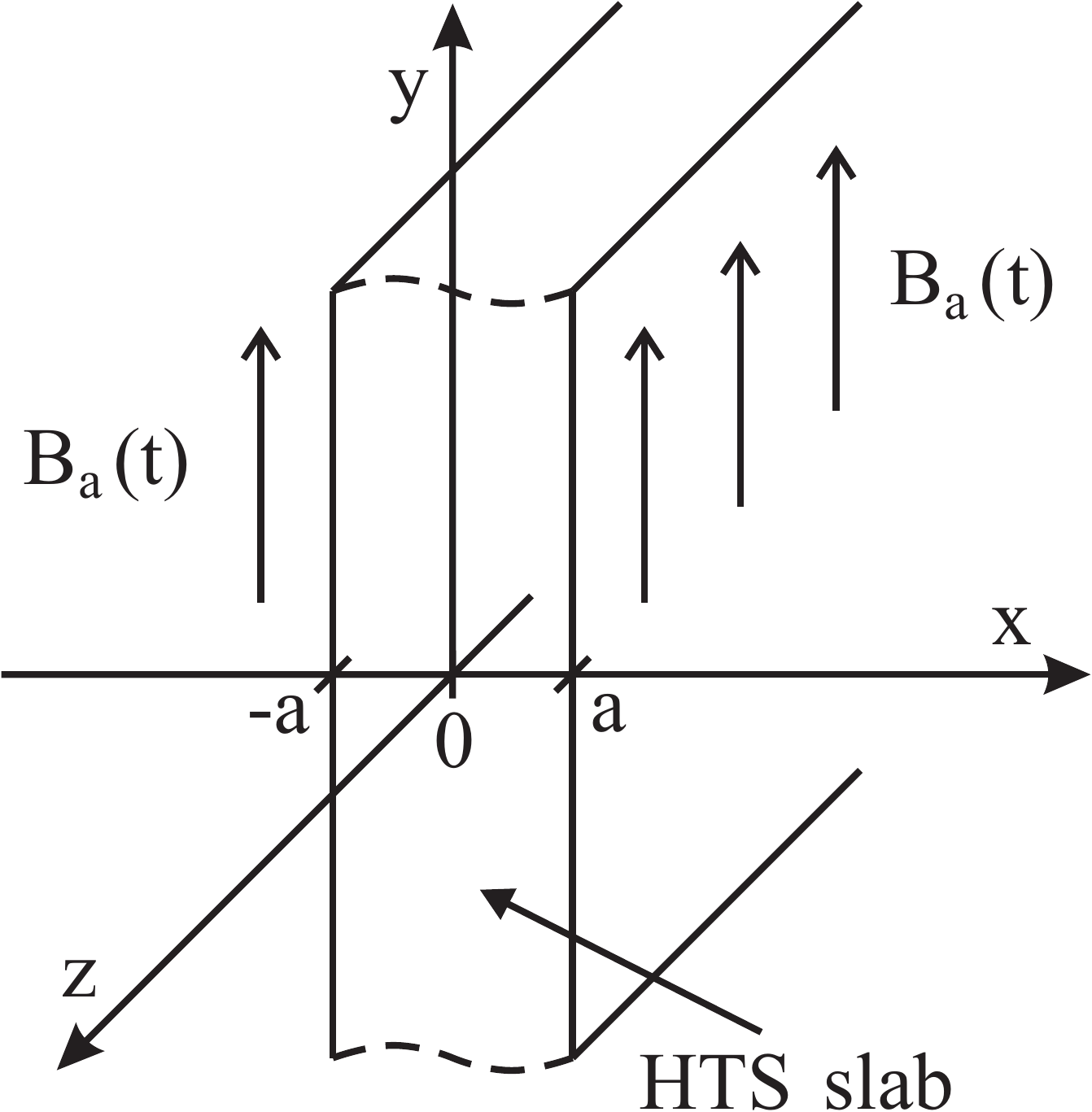}~~~
\includegraphics[width= 9 cm ]{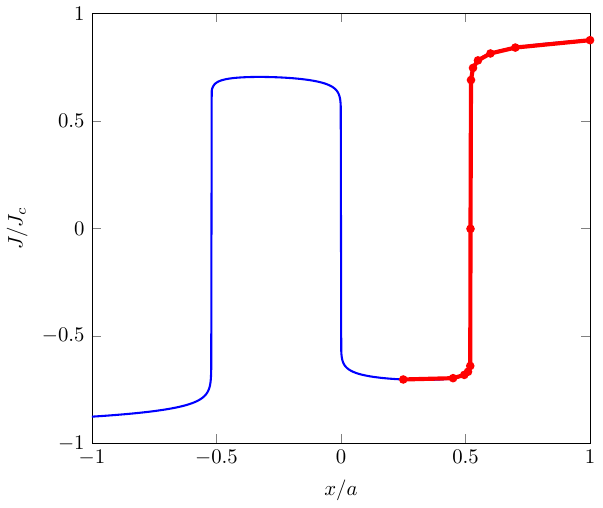}
\caption{\label{fig:slab_FS}{Left: An infinite superconducting slab of width $2a$ subjected to a sinusoidally varying magnetic field parallel to its surface; Right: Current density profile along the $x$-axis, after 2 complete cycles of full penetration of the field. The 11 red line segments in red, when properly located, are sufficient to approximate the current profile, despite the fact that it took approximately 1000 fixed segments in the finite element mesh in order to obtain the same accuracy (in blue) when no adaptivity of the mesh is used.}}
\end{figure}

Unfortunately, current simulators do not allow this optimization: they let the user choose its discretization (e.g. the mesh) in an arbitrary manner. In addition, since the current and flux profiles in superconducting materials are singular and exhibit sharp fronts or cusps that move with time, there is a need to finely mesh all superconducting domains (otherwise one would completely miss the physics), which increases dramatically the number of DOFs to solve for (see figure~\ref{fig:slab_FS} as an example). In addition, due to the strong nonlinearity of the problems, there is a need for very small time steps in order to ensure the convergence of the solver. Therefore, even the simplest problems require many DOFs both in space and time, and the computational time explodes quickly as devices get complicated.

The solution to this problem is to use \emph{adaptivity} in order to refine (or unrefine) the mesh (or the time step) as the solution progresses. The principle of adaptivity is best illustrated from a simple example: the superconducting slab in a parallel magnetic field, as illustrated in figure~\ref{fig:slab_FS}. This figure was obtained by solving the electromagnetic problem on a very fine mesh (2000 DOFs). It shows the solution for the current density $J$ at an arbitrary time instant of a sinusoidal excitation. One can see that we could approximate this solution to a great accuracy using only a few well positioned linear segments. Of course, the position of the vertices defining those line segments has to be adjusted as the front moves, which requires an automated rule to reposition them. For this purpose, one requires \emph{error estimators}, which are the missing tool in today's simulators. Indeed, error estimators evaluate where in space (and/or time) the error is large, which allows deciding where to refine (or unrefine) the mesh. Of course, the computational cost of the error estimator and of the mesh operations has to be as low as possible, but as a general rule, it is quicker (and less memory demanding) to solve a few small problems (the initial mesh and a few refinements) instead of a very big problem that blindly provides good resolution everywhere, including where it is not needed.

Generic adaptive methods to adapt solutions in time exist for systems of ODEs (e.g. see~\cite{Brenan:1996un}), but they are far less popular for PDE, which involve handling mesh adaption. In fluid mechanics and some other fields, those methods have been studied to some extent, but they are in their infancy in computational electromagnetism. As this is a highly technical field, there is a need for the HTS modelling community to increase its interactions with experts in the field, including mathematicians and specialists in numerical analysis. The development of an error estimator dedicated to superconductors would be highly beneficial for this community. Testing of error estimators and adaptive techniques is expected to be easier in home-made codes, as they are more flexible than commercial codes, but hopefully the specialized techniques will make their way to commercial codes as they get mature, making them even more accessible to device designers from the industry.

Finally, it is important to mention that more general progress in parallel computing and numerical solvers for linear systems to solve large problems will help increase computational speed over time. This topic is entirely in the hands of specialists in scientific computing, and any progress on that front cannot really take advantage of the particularities of the problems involving HTS materials. The two approaches are therefore complementary.

\subsection{Availability and easiness-of-use of the simulation tools}

This second path towards improvement is more organizational than technical: the HTS modelling community needs to establish sharing tools and work methodology in order to facilite the progress of everybody and avoid duplication of work or intensive familiarization on simulation tools that will have a short life time. There are many ways to achieve this better collective coordination, for instance: 
\begin{itemize}
\item{establish a web depot for sharing templates of model files, source codes, material models, script files, etc.;}
\item{share codes in an open source format (for code developers), and with proper documentation;}
\item{pursue the effort in developing benchmark models, which are needed to assess objectively the performance of the various tools available;}
\item{increase networking activities among the members of the HTS modelling community: new workshops, summer schools, web forums, etc.}
\end{itemize}

As a first step for sharing codes, models and documentation, the community needs a website, which already exists~\cite{nummodweb}. For instance, one section of this website contains description of benchmarks that were established during the First International Workshop on HTS Modelling~\cite{Lausanne_Workshop}. New benchmarks are expected to be added periodically, as simulations complexity evolve.

Similar sections of the web site can easily be added for templates of model files (especially demo files to be used with commercial software packages), materials data, source codes, etc. The benefits of this increased sharing is obvious: it will allow students, researchers and device designers to jump much more quickly to the core of their analysis, stopping the perpetual reinvention of the wheel every time a new or even a common modelling problem arise.

It is worth mentioning that, despite the fact that a lot of physical model files are available worldwide and could be shared easily, there is still a big gap in macroscopic models, and in particular models that could be implemented directly in a power system or circuit simulators, such as Simulink, EMPT-RV, etc. This kind of models is of prime importance for system engineers, since without them, they cannot assess the impact of a superconducting device in their systems.

Regarding the sharing of source codes, whenever possible, it seems preferable to build on top of existing tools (e.g. Sundials, GetDP, etc.) that are already well documented: that makes more realistic the objective of documenting home-made codes, since the new documentation is then limited to the parts that are specific to HTS modelling. This also prevents reinventing pieces of code that are already mature and well optimized.

As a further organization step, shared codes should eventually become developed in a collective way, with possibilities for different research group to contribute to it. This requires some coordination and the management of a versioning tool, but those are commonly available nowadays (GitHub, etc.), and relatively easy to manage. In addition, the development platform and the scripting tools should be free of charges (for instance, Python is a free scripting language that works on any platform), in order to avoid any legal issue that would prevent contributors to participate in the developments.

Regarding home-made codes developed with the intent to be shared, they should come with examples of applications that are ``ready-to-run'' and documented. Ideally, they should not be more complicated to run than a Maltab script file, so the code could be used by non-expert programmers. This is particularly important, as many device designers from the industry have not been trained as computer scientists, and therefore, they cannot spend time to learn a complex code.

Finally, the real key to collective progress is a good communication between the researchers: this can be achieved through specialized workshops and other communication means, in particular web forums, where device designers and researchers could share their thoughts. No forum on HTS modelling seem to exist at the moment, but creating one would be easy.

\section{Conclusion}

In this paper, we attempted to review globally the current situation of numerical modelling in the context of superconducting device development. Without any doubt, numerical modelling is a key ingredient towards optimized HTS devices, and it is a field in development, especially regarding electromagnetic models. 

The main conclusions of this paper are that, despite the fact that there exist numerous numerical models for accurately simulating the behaviour of HTS devices if one possesses accurate materials models, the computation times of realistic devices are still too long. We also report on a lack of proper benchmarking between existing numerical methods. An even biggest concern is about the availability of models and modelling data themselves, in particular: model file templates, materials data, power system librairies, documented open source simulators, low cost commercial codes, etc.

The unavailability of one or many of these elements considerably slows down the development of HTS devices, and in this paper we suggest some paths toward improvement. One of them is the set up of a website for sharing model files, materials data, modelling experience, benchmarks problems and results, etc. This initiative has already started~\cite{nummodweb} and should improve as modellers get organized in a more structured network.

Finally, regarding the performance of the numerical simulators themselves, further research should be conducted in the field of adaptive methods (both in space and time) and error estimators, in order to automate as much as possible the choice of a minimum but meaningful discretization. Basic efforts in code development should also be pursued collectively between the few groups worldwide who have an expertise in this area, and a good documentation of this work should be done in the short term.

\section*{References}
\bibliographystyle{unsrt}	

\begin{thebibliography}{10}
\expandafter\ifx\csname url\endcsname\relax
  \def\url#1{{\tt #1}}\fi
\expandafter\ifx\csname urlprefix\endcsname\relax\def\urlprefix{URL }\fi
\providecommand{\eprint}[2][]{\url{#2}}

\bibitem{Melhem:2012}
Melhem Z 2012 {\em {High temperature superconductors (HTS) for energy
  applications }\/} (Cambridge, U.K.: Woodhead Publishing)

\bibitem{Matias:PP12}
Matias V and Hammond R~H 2012 {\em Physics Procedia\/} {\bf 36} 1440--1444

\bibitem{Lowther:2013eg}
Lowther D~A 2013 {\em IEEE Transactions on Magnetics\/} {\bf 49} 2375--2380

\bibitem{Campbell:JSNM11}
Campbell A~M 2011 {\em Journal of Superconductivity and Novel Magnetism\/} {\bf
  24} 27--33

\bibitem{Mikitik:TAS13}
Mikitik G~P, Mawatari Y, Wan A~T~S and Sirois F 2013 {\em IEEE Transactions on
  Applied Superconductivity\/} {\bf 23} 8001920

\bibitem{Grilli:TAS14a}
Grilli F, Pardo E, Stenvall A, Nguyen D~N, Yuan W and G{\"o}m{\"o}ry F 2014
  {\em IEEE Transactions on Applied Superconductivity\/} {\bf 24} 8200433

\bibitem{Brambilla:SST08}
Brambilla R, Grilli F, Martini L and Sirois F 2008 {\em Superconductor Science
  and Technology\/} {\bf 21} 105008

\bibitem{Takeuchi:SST11}
Takeuchi K, Amemiya N, Nakamura T, Maruyama O and Ohkuma T 2011 {\em
  Superconductor Science and Technology\/} {\bf 24} 085014

\bibitem{Amemiya:SST14}
Amemiya N, Tsukamoto T, Nii M, Komeda T, Nakamura T and Jiang Z 2014 {\em
  Superconductor Science and Technology\/} {\bf 27} 035007

\bibitem{Chan:TAS10}
Chan W~K, Masson P~J, Luongo C and Schwartz J 2010 {\em IEEE Transactions on
  Applied Superconductivity\/} {\bf 20} 2370--2380

\bibitem{Lausanne_Workshop}
{First International Workshop on Modelling of HTS}
  \urlprefix\url{http://supra.epfl.ch/modellinggroup/Workshop.asp}

\bibitem{Zermeno:JAP13}
Zermeno V~M~R, Abrahamsen A~B, Mijatovic N, Jensen B~B and Soerensen M~P 2013
  {\em Journal of Applied Physics\/} {\bf 114} 173901

\bibitem{Chan:TAS12}
Chan W~K and Schwartz J 2012 {\em IEEE Transactions on Applied
  Superconductivity\/} {\bf 22} 4706010

\bibitem{Clem:SST07}
Clem J~R, Claassen J~H and Mawatari Y 2007 {\em Superconductor Science and
  Technology\/} {\bf 20} 1130--1139

\bibitem{Prigozhin:SST11}
Prigozhin L and Sokolovsky V 2011 {\em Superconductor Science and Technology\/}
  {\bf 24} 075012

\bibitem{Yuan:SST09}
Yuan W, Campbell A~M and Coombs T~A 2009 {\em Superconductor Science and
  Technology\/} {\bf 22} 075028

\bibitem{Campbell:SST07}
Campbell A~M 2007 {\em Superconductor Science and Technology\/} {\bf 20} 292

\bibitem{Meunier:TMAG03}
Meunier G, {Le Floch} Y and {C Gu\'erin} 2003 {\em IEEE Transactions on
  Magnetics\/} {\bf 39} 1729--1732

\bibitem{LeFloch:TMAG03}
{Le Floch} Y, Meunier G, {Gu\'erin} C, Labie P, Brunotte X and Boudaud D 2003
  {\em IEEE Transactions on Magnetics\/} {\bf 39} 1725--1728

\bibitem{Brandt:1996ez}
Brandt E~H 1996 {\em Physical Review B\/} {\bf 54} 4246--4264

\bibitem{Kameni:2010cq}
Kameni A, Mezani S, Sirois F, Netter D, L{\'e}v{\^e}que J and Douine B 2010
  {\em IEEE Transactions on Magnetics\/} {\bf 46} 3445--3448

\bibitem{Bossavit:TMAG94}
Bossavit A 1994 {\em IEEE Transactions on Magnetics\/} {\bf 30} 3363--3366

\bibitem{Prigozhin:JCP96}
Prigozhin L 1996 {\em Journal of Computational Physics\/} {\bf 129} 190--200

\bibitem{Sanchez:PRB01}
Sanchez A and Navau C 2001 {\em Physical Review B\/} {\bf 64} 214506

\bibitem{Pardo:SST07}
Pardo E, G{\"o}m{\"o}ry F, {\v S}ouc J and Ceballos J~M 2007 {\em
  Superconductor Science and Technology\/} {\bf 20} 351--364

\bibitem{Sirois:TAS09}
Sirois F, Roy F and Dutoit B 2009 {\em IEEE Transactions on Applied
  Superconductivity\/} {\bf 19} 3600--3604

\bibitem{Sykulski:TM97}
Sykulski J~K, Stoll R~L, Mahdi A~E and Please C~P 1997 {\em IEEE Transactions
  on Magnetics\/} {\bf 33} 1568--1571

\bibitem{Vinot:TMAG00}
Vinot E, Meunier G and Tixador P 2000 {\em IEEE Transactions on Magnetics\/}
  {\bf 36} 1226--1229

\bibitem{Company:Flux}
Finite-element software package {Flux}.
  http://www.cedrat.com/en/software/flux.html

\bibitem{Grilli:TAS05a}
Grilli F, Stavrev S, {Le Floch} Y, Costa-Bouzo M, Vinot E, Klutsch I, Meunier
  G, Tixador P and Dutoit B 2005 {\em IEEE Transactions on Applied
  Superconductivity\/} {\bf 15} 17--25

\bibitem{Sirois:JPCS08}
Sirois F, Dione M, Roy F, Grilli F and Dutoit B 2008 {\em Journal of Physics:
  Conference Series\/} {\bf 97} 012030

\bibitem{Brenan:1996un}
Brenan K~E, Campbell S~L and Petzold L~R 1996 {\em {Numerical solution of
  initial-value problems in differential-algebraic equations }\/}
  (Philadelphia, PA: SIAM)

\bibitem{Lahtinen:2012bp}
Lahtinen V, Lyly M, Stenvall A and Tarhasaari T 2012 {\em Superconductor
  Science and Technology\/} {\bf 25} 115001

\bibitem{Website:TEAM}
{International Compumag Society: TEAM}
  \urlprefix\url{http://www.compumag.org/jsite/team.html}

\bibitem{Pecher:2003wm}
Pecher R, McCulloch M~D, Chapman S~J, Prigozhin L and Elliott C~M 2003
  {3D-modelling of bulk type-II superconductors using unconstrained
  H-formulation} {\em 11th European Conference on Applied Superconductivity
  (EUCAS)\/} (Sorrento, Italy, Sept. 14-18)

\bibitem{Pellikka:2013ha}
Pellikka M, Suuriniemi S, Kettunen L and Geuzaine C 2013 {\em SIAM Journal on
  Scientific Computing\/} {\bf 35} B1195--B1214

\bibitem{Sirois:2009tn}
Sirois F, Coulombe J and Bernier A 2009 {\em IEEE Transactions on Applied
  Superconductivity\/} {\bf 19} 3585--3590

\bibitem{Kim:PRL62}
Kim Y, Hempstead C and Strnad A 1962 {\em Physical Review Letters\/} {\bf 9}
  306--309

\bibitem{Stavrev:TAS02}
Stavrev S, Dutoit B and Nibbio N 2002 {\em IEEE Transactions on Applied
  Superconductivity\/} {\bf 3} 1857--1865

\bibitem{Pardo:SST11}
Pardo E, Vojen{\v c}iak M, G{\"o}m{\"o}ry F and {\v S}ouc J 2011 {\em
  Superconductor Science and Technology\/} {\bf 24} 065007

\bibitem{Grasso:PhysC95}
Grasso G, Hensel B, Jeremie A and Fl{\"u}kiger R 1995 {\em Physica C\/} {\bf
  241}

\bibitem{Amemiya:PhysC06b}
Amemiya N, Maruyama O, Mori M, Kashima N, Watanabe T, Nagaya S and Shiohara Y
  2006 {\em Physica C\/} {\bf 445-448} 712--716

\bibitem{Grilli:TAS07a}
Grilli F, Brambilla R and Martini L 2007 {\em IEEE Transactions on Applied
  Superconductivity\/} {\bf 17} 3155--3158

\bibitem{Solovyov:SST13}
Solovyov M, Pardo E, Souc J, {G\"om\"ory} F, Skarba M, Konopka P, {Pekar\v
  c\'ikov\'a} M and Janovec J 2013 {\em Superconductor Science and
  Technology\/} {\bf 26} 115013

\bibitem{Grilli:Cryo13}
Grilli F, Brambilla R, Sirois F, Stenvall A and Memiaghe S 2013 {\em
  Cryogenics\/} {\bf 53} 142--147

\bibitem{Lacroix:SST14}
Lacroix C and Sirois F 2014 {\em Superconductor Science and Technology\/} {\bf
  27} 035003

\bibitem{Nguyen:SST10}
Nguyen D~N, Ashworth S~P, Willis J~O, Sirois F and Grilli F 2010 {\em
  Superconductor Science and Technology\/} {\bf 23} 025001

\bibitem{Pardo:SST12b}
Pardo E, {\v S}ouc J and Kov{\'a}{\v c} J 2012 {\em Superconductor Science and
  Technology\/} {\bf 25} 035003

\bibitem{Grilli:TAS14b}
Grilli F, Zermeno V~M~R, Pardo E, Vojenciak M, Brand J, Kario A and Goldacker W
  2014 {\em IEEE Transactions on Applied Superconductivity\/} {\bf 24} 4801005

\bibitem{BadiaMajos:2012dt}
Bad{\'\i}a-Maj{\'o}s A and L{\'o}pez C 2012 {\em Superconductor Science and
  Technology\/} {\bf 25} 104004 (16 pp.)

\bibitem{Grilli:TAS14c}
Grilli F, Sirois F, Zerme{\~n}o V~M~R and Vojen\v{c}iak M 2014 {\em IEEE
  Transactions on Applied Superconductivity\/}

\bibitem{nummodweb}
{HTS Modelling Workgroup} \urlprefix\url{http://www.htsmodelling.com/}

\end{thebibliography}

\providecommand{\newblock}{}

\end{document}